\documentclass[apjl]{emulateapj}

\usepackage{graphicx,color,amsmath,amssymb,amsfonts}
\usepackage[usenames,dvipsnames]{xcolor}

\usepackage{hyperref}
\hypersetup{
    colorlinks = true,
    citecolor = {MidnightBlue},
    linkcolor = {BrickRed},
    urlcolor = {BrickRed}
}

\newcommand{\be}{\begin{equation}}
\newcommand{\ee}{\end{equation}}
\newcommand{\bea}{\begin{eqnarray}}
\newcommand{\eea}{\end{eqnarray}}
\newcommand{\bfig}{\begin{figure}}
\newcommand{\efig}{\end{figure}}
\newcommand{\nn}{\nonumber}

\def\fnl{{f_{\rm NL}}}
\def\l{\left(}
\def\r{\right)}

\def\z{(z)}
\def\fsky{f_{\rm sky}}
\def\lm{{\ell_{\rm min}}}
\def\lM{{\ell_{\rm max}}}
\def\mH{{\mathcal H}}
\def\p{\partial}

\def\cs{{\sc camb}$\_$sources }
\def\D{\Delta}
\def\euclid{Euclid}

\def\odm{\Omega_{\rm DM}}
\def\mag{\mathcal{Q}}

\def\d{\mathrm{d}}

\def\fgr{{f_{\rm GR}}}

\def\s{{\sigma}}

\begin{document}

\shorttitle{Hunting down horizon-scale effects with multi-wavelength surveys}
\shortauthors{Jos\'e Fonseca \textit{et al.}}

\title{Hunting down horizon-scale effects with multi-wavelength surveys}

\author{Jos\'e Fonseca,\altaffilmark{1}$^*$ Stefano Camera,\altaffilmark{2} M\'ario G. Santos\altaffilmark{1}$^,$\altaffilmark{3}$^,$\altaffilmark{4}, Roy Maartens\altaffilmark{1}$^,$\altaffilmark{5}}
\affiliation{$^1$ Physics Department, University of the Western Cape, Cape Town 7535, South Africa; {\color{blue} josecarlos.s.fonseca@gmail.com}}
\affiliation{$^2$ Jodrell Bank Centre for Astrophysics, The University of Manchester, Manchester M13 9PL, UK}
\affiliation{$^3$ SKA South Africa, The Park, Cape Town 7405, South Africa}
\affiliation{$^4$ CENTRA, Instituto Superior T\'ecnico, Universidade de Lisboa, Lisboa, 1049-001, Portugal}
\affiliation{$^5$ Institute of Cosmology \& Gravitation, University of Portsmouth, Portsmouth PO1 3FX, UK}

\begin{abstract}
Next-generation cosmological surveys will probe ever larger volumes of the Universe, including the largest scales, near and beyond the horizon. On these scales, the galaxy power spectrum carries signatures of local primordial non-Gaussianity (PNG) and horizon-scale general relativistic (GR) effects. 
However, cosmic variance limits the detection of horizon-scale effects. Combining different surveys via the multi-tracer method allows us to reduce the effect down cosmic variance.
This method benefits from large bias differences between two tracers of the underlying dark matter distribution, which suggests a multi-wavelength combination of large volume surveys that are planned on a similar timescale. We show that the combination of two contemporaneous surveys, a large neutral hydrogen intensity mapping survey in SKA Phase\,1 and a \euclid-like photometric survey, will provide unprecedented constraints on PNG as well as detection of the GR effects. We forecast that the error on local PNG will break through the cosmic variance limit on cosmic microwave background surveys and achieve $\s(\fnl)\simeq1.4-0.5$, depending on assumed priors, bias, and sky coverage. GR effects are more robust to changes in the assumed fiducial model, and we forecast that they can be detected with a signal-to-noise of about $14$.

\end{abstract}

\maketitle

\section{Introduction}

Upcoming cosmological surveys will probe larger volumes of the Universe, opening new windows to studying cosmological effects on horizon scales \citep[see e.g.][]{Yoo:2012se,Alonso:2015uua,Camera:2015fsa,Raccanelli:2015vla}. These effects include primordial non-Gaussianity (PNG) and general relativistic (GR) horizon-scale effects in the observed power spectrum. 

PNG is a key discriminator between different classes of inflation models. Local-type PNG  (characterised by the parameter $\fnl$) leaves a frozen imprint on horizon-scale power, allowing us to probe the primordial Universe via the cosmic microwave background (CMB) and large-scale structure surveys. The Planck constraint \citep{Ade:2015ava}, $\sigma(\fnl)\simeq 6.5$ (using the large-scale structure convention), is far stronger than those from current galaxy surveys, but is close to the maximum achievable with CMB experiments, which can only rule out inflation models with relatively large PNG. 

Local PNG induces a scale-dependent correction to the bias of any dark matter tracer \citep{Dalal:2007cu,Matarrese:2008nc}. This scale dependence can be probed through the two-point correlation function of the tracer on very large scales, allowing next-generation surveys to significantly improve upon the CMB constraints \citep[see, e.g.][]{Giannantonio:2011ya,Camera:2013kpa,Camera:2014bwa}.

At this level of sensitivity, neglecting GR horizon-scale effects would bias results. Moreover, they might hint at something new if GR breaks down on these scales. They arise via lightcone observations of dark matter tracers such as the number counts of galaxies \citep{Yoo:2010ni,Challinor:2011bk,Bonvin:2011bg} or maps of intensity \citep[e.g. the integrated 21cm signal from neutral hydrogen (\textsc{Hi}) galaxies;][]{Hall:2012wd}, including Doppler, Sachs--Wolfe, integrated Sachs--Wolfe and time-delay-type terms. The lensing contribution to the clustering power, mediated by magnification bias, can also be significant on horizon scales \citep{Alonso:2015uua,Montanari:2015rga}. 

Cosmic variance becomes a serious obstacle for horizon-scale measurements where PNG and GR signals are strongest. Forecasts for next-generation surveys show that GR effects will not be detectable using a single tracer and PNG detection is limited to $\sigma(\fnl) > 1$ \citep{Alonso:2015uua,Raccanelli:2015vla}. This calls for the multi-tracer technique (MT) to beat down cosmic variance \citep{McDonald:2008sh,Seljak:2008xr}. 

MT has been used to explore improvements in the measurement of $\fnl$  \citep[see e.g.][]{McDonald:2008sh,Hamaus:2011dq,Abramo:2013awa,Ferramacho:2014pua, Yamauchi:2014ioa}. In these works, the lensing and GR contributions to clustering power were ignored. While this may have little effect on $\s(\fnl)$, it can significantly bias the best-fit value extracted from the data \citep{Namikawa:2011yr,Camera:2014sba}. MT has also been used to forecast detectability of GR effects by \cite{Yoo:2012se}, but neglecting the lensing contribution and the integrated GR effects. Here we include all lensing and GR effects without making any flat-sky approximation.

The MT technique opens a new observational window into probing large-scale signatures in the Universe. In addition to reducing cosmic variance, it also cancels the individual systematics of the two experiments and removes foreground residuals. We show here that MT is a game-changer in the way we design surveys to probe these scales, as volume is no longer the ultimate goal and noise reduction becomes a priority again.


\section{The Multi-Tracer Technique}

The theoretical observed fluctuations for a given dark matter tracer $A$ can be written in Fourier space and Newtonian gauge in the form:
\bea
\Delta^A &&=\delta\Bigg\{b_{\rm G}^A+\Delta b^A+f\frac{k_\|^2}{k^2}+E({\cal Q}^A-1)\frac{k_\perp^2}{k^2}\nn\\
&&+\left[F-{\rm i}f\left(b_{e}^A-2{\cal Q}^A+\frac{2{\cal Q}^A}{\mH\chi}\right)\frac{k_\|}{k}\right] \frac{\mH}{k}\label{eq:Delta_A} \\
&&+\left[G+fb_{e}^A+I{\cal Q}^A+J\left(b_{e}^A-2{\cal Q}^A+\frac{2{\cal Q}^A}{\mH\chi}\right)\right]{\mH^2\over k^2}\Bigg\}\nn,
\eea
where $\mathbf k=(\mathbf k_\perp,k_\|)$ and ${\cal H}$ is the conformal Hubble parameter. The first line contains the RSD and lensing terms, while the next two lines constitute the horizon-scale GR terms. The density contrast $\delta$ is in the comoving-synchronous gauge in order to consistently define the bias on large scales, and $f$ is the growth rate. The correction to the Gaussian bias $b_{\rm G}$ due to local PNG is given by $\D b(z,k)= 3\fnl [ b_{\rm G}(z) -1]\Omega_m H_0^2 \delta_c/[D(z) T(k)  k^2]$. Here $\delta_c\simeq 1.69$ is the critical matter density contrast for spherical collapse, $T\l k\r$ is the transfer function (normalised to 1 on large scales), and $D\l z \r$ is the growth factor (normalised to 1 at $z=0$). 
The magnification bias is $\mag(z,\mathcal F_{\ast})\equiv[-\p\ln N_{\rm s}/\p\ln\mathcal F]_{\mathcal F=\mathcal F_\ast}$, where $N_{\rm s}(z,\mathcal F>\mathcal F_{\ast})$ is the background number density of sources at redshift $z$ with flux $\mathcal F$ above the detection threshold $\mathcal F_{\ast}$. The evolution bias is $b_e(z) = -\p\ln[(1+z)^{-3}N_{\rm s}]/\p\ln (1+z)$. For details on the tracer-independent background functions $E,F,G,I$ and $J$ see \citet{Challinor:2011bk}.

We see that PNG grows as $k^{-2}$, while GR terms grow as ${\cal H}/k$ or ${\cal H}^2/k^2$. The difference in the scale dependence and amplitude of the different terms allows GR corrections to be distinguished from PNG. More importantly, cosmic variance uncertainties come from $\delta$, since it is a single realisation of the underlying probability distribution. The MT technique relies on the fact that the ratio of different $\Delta$'s is independent of cosmic variance, since both tracers are linear in $\delta$.

The estimator we use is the sky map itself in the form of the $a_{\ell m}$. Assuming that the  $a_{\ell m}$ are Gaussian, all the information will be encoded in the angular power spectrum, $\left < a_{\ell m} a^*_{\ell' m'}\right > = \delta_{\ell \ell'}\delta_{m m'} C_\ell$. Using a Gaussian likelihood for the $a_{\ell m}$, the corresponding Fisher matrix will then be enough to account for the MT effects. 
Extending the single tracer case \citep{Challinor:2011bk} to multi-tracers, the angular power spectrum is given by
\be \label{cl}
C^{AB}_\ell\l z_i,z_j\r=4\pi\int\!\!\d \ln k\,\D_\ell^{W_A}\l z_i,k\r \D_\ell^{W_B}\l z_j,k\r \mathcal P_\zeta\l k\r \!.
\ee
Here, $z_i$ are the redshift bin centres and $\mathcal P_\zeta$ is the dimensionless power spectrum of the primordial curvature perturbation. The measurable transfer function in the bin is
\be \label{transf}
\D_\ell^{W_A}\l z_i,k\r=\int\!\! \d z\,p^A(z)W(z_i,z)\D_\ell^A(z,k),
\ee
where $p^A(z)$ is the redshift distribution function of tracer $A$. 

Equation (\ref{eq:Delta_A}) gives the theoretical transfer function $\D^{ A}_\ell(z,k)$, while $W(z_i,z)$ is the window function centered on $z_i$, namely the probability distribution function of a source to be inside the $i$th bin. The product $p^A(z)W(z_i,z)$ is the effective tracer's redshift distribution function inside the bin, normalised so that $\int\d z\,p^A(z)W(z_i,z)=1$ for all $z_i$.

\section{Multi-wavelength surveys}\label{sec:surveys}
 
In order to optimally exploit the MT method, we look for two surveys with significant difference in bias. We focus on the \textsc{Hi} intensity mapping (IM) survey that will be performed with SKA phase 1 in `single' dish mode \citep{Maartens:2015mra,Santos:2015bsa} together with a Euclid-type galaxy survey \citep{Laureijs:2011gra,Amendola:2012ys}. We opt for the planned photometric survey because it will detect a larger number of galaxies than the spectroscopic option. We also consider variations to this survey with different noise and sky coverage, similar to a second-generation galaxy survey such as the Large Synoptic Survey Telescope (LSST; \citealt{Abate:2012za,Bacon:2015dqe}). 

\paragraph*{\textsc{Hi} IM experiment.} In \textsc{Hi} IM, all galaxies with \textsc{Hi} contribute to the signal. We compute the Gaussian \textsc{Hi} bias, $b^{\rm \textsc{Hi}}_{\rm G}\z$, weighting the halo bias with the \textsc{Hi} content in the dark matter halos \citep{Santos:2015bsa}. The number of observed sources is independent of the flux limit but the fractional temperature perturbation is equal to Eq. (\ref{eq:Delta_A}) with $\mag^{\rm \textsc{Hi}}=1$ \citep{Hall:2012wd}. The signal's redshift distribution follows the \textsc{Hi} temperature, $p^{\rm \textsc{Hi}}(z)\propto T_{\rm \textsc{Hi}}(z)$, which we fit using the results of \citet{Santos:2015bsa}. For \textsc{Hi} IM, $N_{\rm s}/(1+z)^3=n_{\rm \textsc{Hi}}\z\propto T_{\rm \textsc{Hi}}(z) H\z /(1+z)^2$ is the comoving density of \textsc{Hi} atoms \citep{Hall:2012wd}, which is used to compute $b_e^{\rm \textsc{Hi}}$. The noise angular power spectrum in the $i$th bin of frequency width $\D\nu_i$ for an experiment  with $N_d$ collecting dishes, total observation time $t_{\rm tot}$ and observed fraction of the sky $\fsky$, is given by
\be
\mathcal N_{\rm \textsc{Hi}}^{ij}=\frac{4\pi \fsky T^2_{\rm sys}}{2N_d t_{\rm tot}\Delta\nu_i}\delta^{ij},\label{eq:N_HI}
\ee
where $T_{\rm sys}=25+60(300\,{\rm MHz}/\nu)^{2.55}\,\rm K$ is the system temperature. For  SKA1, we assume $N_d t_{\rm tot}=2\times10^6\,\rm hr$ and  $\fsky=0.72$. For \textsc{Hi} IM one can neglect shot noise \citep{2011ApJ...740L..20G}.

\paragraph*{Photometric galaxy survey (PG).}
The bias, magnification bias, and photometric galaxies' redshift distribution that we adopt are \citep{Amendola:2012ys,Raccanelli:2015vla}
\begin{align}
b^{\rm PG}_{\rm G}\z&=\sqrt{1+z},\label{eq:Euclid-bias}\\
{\cal Q}^{\rm PG}(z)&=0.2985 + 0.5305 z -0.1678 z^2 + 0.2578z^3,\\
p^{\rm PG}(z)&\propto z^2 \exp\big[-(1.412z/0.9)^{3/2}\big],
\end{align}
where the proportionality in the last equation is set by the total number of galaxies detected. For a galaxy survey the noise angular power spectrum is dominated by shot noise, i.e.,
\be
\mathcal N_{\rm PG}^{ij}=\frac{\delta^{ij}}{N_{\rm PG}^i},
\ee
where $N_{\rm PG}^i$ is the number of galaxies per steradian in the $i$th bin. For the scatter between the photometric redshift estimate and the true redshift, we use $\s_{\rm ph}(z)=0.05(1+z)$ \citep{Ma:2005rc}.

We consider three observational scenarios: ({\bf\emph{i}})~\euclid-like survey detecting 30 galaxies per square arcminute and covering 15,000 $\rm deg^2$, with 50\% overlap with the SKA1 \textsc{Hi} experiment; ({\bf\emph{ii}})~the same case but with 100\% overlap; ({\bf\emph{iii}})~a more futuristic LSST-like survey detecting 40 galaxies per square arcminute and covering the whole SKA1 sky. Summarising,
\begin{align}\label{pgscen}
\fsky^{(i)}&=\frac{1}{2}\fsky^{(ii)}=\frac{1}{4}\fsky^{(iii)}=0.18 \\
N_{\rm PG}^{(i)}&=N_{\rm PG}^{(ii)}=\frac{3}{4}N_{\rm PG}^{(iii)}=30\,\mathrm{arcmin}^{-2}.
\end{align}

When referring to a single tracer, we use its value of $\fsky$, while in the case of MT for $(i)$ and $(ii)$ only the overlapping sky fraction is considered. Note that for $(i)$ and $(ii)$, we allow the SKA survey to be optimised for the smaller sky area, while maintaining the same total observation time, which will make the noise decrease (Eq.~\eqref{eq:N_HI}). The same cannot be done for the galaxy survey, since its sky coverage is assumed to be already fixed.

\begin{figure}
\centering
\includegraphics[width=\columnwidth]{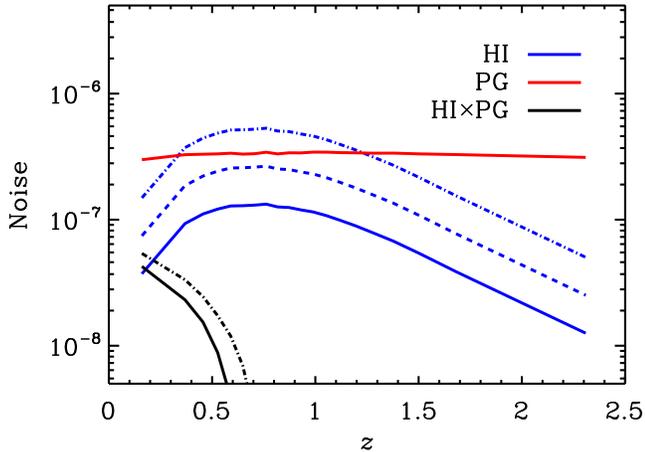}
\caption{Dimensionless noise and cross-noise. The solid lines correspond to $\fsky=0.18$, dashed to $\fsky=0.36$ and dotted-dashed to $0.72$, i.e., the scenarios in Eq.~\eqref{pgscen}. Note that $(i)$ and $(ii)$ have the same cross-noise since $N_{\rm PG}^{(i)}=N_{\rm PG}^{(ii)}$.}\label{Xnoise}
\end{figure}

\paragraph*{\textsc{Hi}-PG Cross-Noise.}
We also need to take into account the possible shot noise cross-power spectrum in each bin. This is due to an overlap in the halo mass range which the tracers probe. Even if this is small, it might be important for the MT as this is the only noise showing up in the cross-correlation between tracers.
The cross-noise is given by
\be
\mathcal N_{\rm \textsc{Hi},PG}^{ij}= \frac{\delta^{ij}T_{\rm \textsc{Hi}}(z_i)}{\rho_{\rm \textsc{Hi}}(z_i)N_{\rm PG}^i}\int\!\!{\rm dM}~\frac{\d N}{\d M} M_{\rm \textsc{Hi}}\l M\r \Theta\l M\r,
\ee
where $\d N/\d M$ is the halo mass function, $M_{\rm \textsc{Hi}}$ is the mass of \textsc{Hi} in a halo of mass $M$, and $\rho_{\rm \textsc{Hi}}(z_i)$ is the \textsc{Hi} density. If the halo masses probed by the two surveys overlap, then $\Theta(M)=1$; otherwise it is zero. For further details on the halo mass range for the SKA \textsc{Hi} IM survey, see \cite{Santos:2015bsa}. The mass range for the photo-$z$ survey is found by matching the number of galaxies in the bin given by the halo mass function with the number given by the redshift distribution of photometric galaxies.

Figure~\ref{Xnoise}  shows the (dimensionless) noise and cross-noise for the chosen binning. The cross-noise is only different from zero at low redshifts when some halos will be detected by both surveys. While \textsc{Hi} IM is sensitive to the low halo mass galaxies, galaxy surveys are more sensitive to massive halos. For the bins with cross-noise, this is at least two orders of magnitude lower than the cross-angular power spectrum at the large-angle multipoles. This might not be the case for more sensitive future surveys (like LSST), which may detect lower mass halos.

\section{Forecasting methodology}
We perform a Fisher analysis \citep{Tegmark:1996bz} for a set of $64$ parameters
\be
  \vartheta_\alpha=\{\fgr,\fnl,\ln A_s,\ln\odm,\ln b_i^A,\ln\mag_i\}, 
\ee  
with $b_i^A\equiv b_{\rm G}^A(z_i)$ (40 parameters) and $\mag_i\equiv\mag^{\rm PG}(z_i)$ (20 parameters). We consider a bias parameter per bin, since due to the high precision measurements we are achieving with the MT technique, using a smaller number of bins with interpolation of parameters might impose a rather strong prior.

We assume a fiducial concordance cosmology with $H_0=67.74\,\mathrm{km/s/Mpc}$, cold dark matter fraction $\odm=0.26$, baryon fraction $\Omega_{\rm b}=0.05$, amplitude of primordial scalar perturbations $A_{\rm s}=2.142\times 10^{-9}$ and $\fnl=0$. The fiducial values for the galaxy bias and magnification bias are set by the survey specifications in \S~\ref{sec:surveys}. 
The new parameter $\fgr$, used as a quantifier of the measurability of GR effects, is defined by
\be
C_\ell=C_\ell^{\delta+{\rm RSD}+{\rm lens}}+\fgr C_\ell^{\rm GR},
\ee
where we take $\fgr=1$ as the fiducial value and $C_\ell^{\delta+{\rm RSD}+{\rm lens}}$ accounts for auto- and cross-correlations between the density, RSD, and lensing terms.
The term $C_\ell^{\rm GR}$ includes all correlations that contain a GR horizon-scale term. 
We chose this definition in order to compute $\p C_\ell/\p\fgr$ analytically.
We include all effects in Eq. (\ref{eq:Delta_A}) at the same time in order not to bias the accuracy on parameter reconstruction \citep[see, e.g.][]{Namikawa:2011yr, Camera:2014dia, Camera:2014sba}.

\begin{table*}
\caption{\label{tab:table}Marginal errors on $\fgr$ and $\fnl$ for a \euclid-like photo-$z$ survey, \textsc{Hi} intensity mapping with SKA1, and their combined MT analysis.}
\begin{ruledtabular}
\begin{tabular}{lccccccccccccc}
&\multicolumn{6}{c}{$\s(\fgr)$}&&\multicolumn{6}{c}{$\s(\fnl)$}\\
\cline{2-7}\cline{9-14}
&\multicolumn{2}{c}{\textsc{Hi}}&\multicolumn{2}{c}{PG}&\multicolumn{2}{c}{MT}&&\multicolumn{2}{c}{\textsc{Hi}}&\multicolumn{2}{c}{PG}&\multicolumn{2}{c}{MT}\\
$\lM$&60&300&60&300&60&300&&60&300&60&300&60&300\\
\hline
without $\{\ln b_i^A,\ln\mag_i\}$ &1.36&1.33&1.58&1.55&0.075&0.072&&4.57&4.31&5.34&5.13&1.23&1.12\\
with $\{\ln b_i^A,\ln\mag_i\}$ &1.39&--&1.90&--&0.079&--&&5.24&--&6.02&--&1.37&--\\
with $\{\ln b_i^A,\ln\mag_i\}$ + 5\% prior &1.38&--&1.68&--&0.076&--&&5.62&--&5.53&--&1.36&--\\
\end{tabular}
\end{ruledtabular}
\end{table*}

We use 20 redshift bins in the range $0<z<3$, with variable size such that approximately the same number of photo-$z$ galaxies resides in each bin. Then, we adopt exactly the same redshift binning for SKA1 \textsc{Hi} IM, so that there is a complete overlap between the two tracers. We can do so thanks to the high resolution of an IM experiment, which allows us to tune the frequency windows. For the MT covariance matrix, we follow \citet{Ferramacho:2014pua}. We focus on large scales, neglecting beam effects from limited angular resolution, which should be negligible for small $\ell$.

The angular power spectra for MT are computed for a Gaussian window function using a modified version of the publicly available \cs code \citep{Challinor:2011bk}. The code was changed to include the tracers' redshift distribution and to compute $b_e^{\rm \textsc{Hi}}$.

We improve the Fisher matrix analysis to ensure numerical stability. Derivatives with respect to $\fgr$ and $\ln b_i^A$ are analytical, and numerical differentiation with respect to other parameters is done using a five-point stencil method. We use the logarithm of the parameter making the Fisher matrix less prone to numerical issues.

The high dimensionality of the tomographic matrices and the large number of nuisance parameters (60) require utmost control on the matrix operations. Therefore, we perform matrix inversion via ``inverse diagonalisation": given a square matrix, its inverse is $\mathbf A^{-1}=\mathbf U \boldsymbol\Lambda^{-1}\mathbf U^{-1}$, where $\mathbf U$ and $\boldsymbol\Lambda$ are respectively the matrices of the eigenvectors and eigenvalues of $\mathbf A$. Thus, $\boldsymbol\Lambda$ is diagonal by construction and its inversion is trivial. This also helps in removing degeneracies in the Fisher matrix. Indeed, when marginalising over the set of nuisance parameters, if one or more eigenvalues (nearly) vanish, then this degeneracy does not propagate into the cosmological parameters of interest \citep{Camera:2012ez}.

\section{Discussion of Results}

Unless otherwise stated, we assume configuration $(i)$ for the photo-$z$ survey, which is the more near-term scenario. For the Fisher matrix, we set the maximum available angular scale, $\lm$, to 2, and consider two different minimum angular scales, $\lM=60$ and 300. Table~\ref{tab:table} shows the 1$\s$ marginal error on $\fgr$ and $\fnl$ for the different tracer configurations, $\lM$ and 60 nuisance parameters. 
Given the high dimensionality of both the $C_\ell$ tomographic matrix and the Fisher matrix, as well as the various implementations ensuring numerical stability of the matrix operations, the computation of the Fisher matrix can become unwieldy as $\lM$ increases. Therefore in Table~\ref{tab:table} we present results for $\lM=300$ for the cosmological parameter set, and show the trend due to the inclusion of the nuisance parameters for $\lM=60$ only.
The big improvement of MT over the single tracer is apparent: for $\fnl$ we get constraints $\sim5$ times tighter, and for $\fgr$ the improvement is even more impressive as the bound shrinks by a factor $>20$. Moreover, MT is more robust when we allow for full uncertainty on the bias-related nuisance parameters, as can be seen in the impact of a 5\% prior on the nuisance parameters. 

The use of two different maximum angular multipoles is done not for the sake of a conservative versus optimistic comparison: both $\lM$'s are well within the linear r\'egime and the inclusion, besides $A_{\rm s}$, of $\odm$ and the nuisance parameters ensures that we do not overestimate the constraining power on $\fnl$ or $\fgr$, even when pushing to small scales. Instead, we want to understand to what extent smaller scales contribute to the signal of PNG or GR effects, both of which are strongest on ultra-large scales. This also enables us to monitor the impact of noise. Noise usually dominates on small scales and is negligible in the cosmic variance limited r\'egime, but \citet{Seljak:2008xr} and \citet{Ferramacho:2014pua} suggested that the more MT is effective in `removing' cosmic variance, the larger the scales at which noise starts becoming relevant.

\begin{figure}
\centering
\includegraphics[width=\columnwidth]{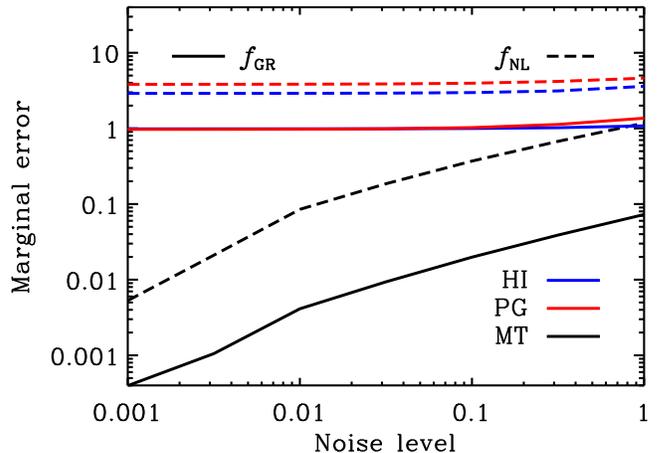}
\caption{Marginal 1$\s$ error on $\fgr$ (solid) and $\fnl$ (dashed) versus noise level for SKA1 intensity mapping (blue), \euclid-like photo-$z$ galaxies (red) and MT (magenta) for $\lM=100$.}\label{fig:sigmaVSnoise}
\end{figure}

In Fig.~\ref{fig:sigmaVSnoise} we show the forecast marginal errors on $\fgr$ (solid) and $\fnl$ (dashed) as a function of the noise level. We multiply the noise of both IM and galaxy number counts by a fudge factor and let it vary from 0 to 1, where 0 means a noiseless experiment and 1 is the real setting. As we remove noise, single tracers soon reach the cosmic variance limited plateau, while MT keeps improving. So, as cosmic variance fades, the more important the signal-to-noise ratio becomes. 

\citet{Seljak:2009af} proposed a mass-dependent weighting of the detected sources, which can considerably suppress the stochasticity between halos and dark matter, thus reducing the shot noise contribution. By doing so, they showed that it will be possible for a next-generation \euclid-like survey to reduce the Poisson noise even by 30\%. From Fig.~\ref{fig:sigmaVSnoise}, the resulting improvement appears very clear \citep[see also][]{Hamaus:2011dq}.

\begin{figure}
\centering
\includegraphics[width=\columnwidth]{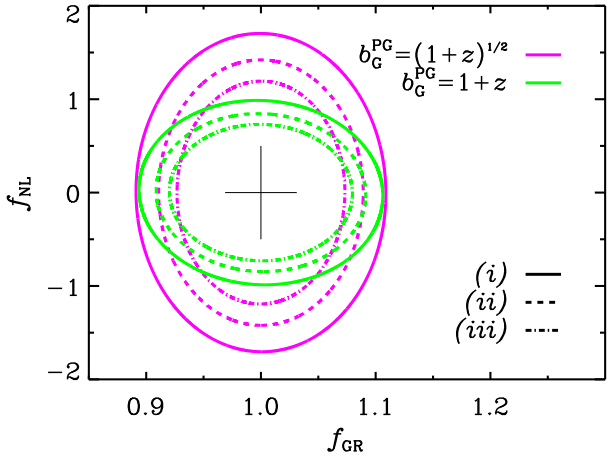}
\caption{Joint 1$\s$ marginal error contours for $\lM=300$ without $\{\ln b_i^A,\ln\mag_i\}$ for the two photo-$z$ Gaussian bias models and the three survey scenarios.}\label{fig:ellipses}
\end{figure}

The impact of $\fsky$ and $b^{\rm PG}_{\rm G}$ is studied by performing the same analysis as before, but comparing the three photo-$z$ scenarios $(i)$, $(ii)$, and $(iii)$ (\S~\ref{sec:surveys}) and changing from the bias of Eq.~\eqref{eq:Euclid-bias} to a higher bias $b^{\rm PG}_{\rm G}=1+z$. In Fig.~\ref{fig:ellipses} we present the corresponding forecast joint 1$\s$ marginal error contours in the $(\fgr,\fnl)$ plane. The collapse from the outermost to the middle ellipses is simply caused by the doubling of the surveyed sky areas, but the innermost contours are also affected by a reduced photometric galaxy shot noise as the number density grows by 30\%. Also, we remind the reader that the \textsc{Hi} IM noise is linearly dependent on $\fsky$. Summarising, in Table~\ref{tab:table-MT} we quote the forecast marginal errors on the measurement of GR effects and PNG for the three photo-$z$ scenarios and two biases. Results are shown for $\lM=300$ and no nuisance parameters.

\begin{table}
\caption{\label{tab:table-MT}Marginal errors from the MT analysis for the three photo-$z$ scenarios with Gaussian bias $\sqrt{1+z}$ (or $1+z$) and $\ell_{\rm max}=300$.}
\begin{ruledtabular}
\begin{tabular}{lcccc}
&\multicolumn{2}{c}{$\s(\fgr)$}&\multicolumn{2}{c}{$\s(\fnl)$}\\
\hline
$(i)$&0.071&(0.070)&1.12&(0.65)\\
$(ii)$&0.059&(0.060)&0.94&(0.56)\\
$(iii)$&0.048&(0.053)&0.79&(0.48)
\end{tabular}
\end{ruledtabular}
\end{table}

To conclude, the MT technique will allow synergies between SKA and Euclid to provide game-changing measurements on horizon scales. We have shown that this can break through the PNG barrier of $\s(\fnl)=1$ and make the first ever detections of the GR effects. Moreover, our analysis shows that with this new method, we need to rethink the way large-scale surveys are being designed. Ultra-large volumes are no longer the ultimate goal, as we can cancel cosmic variance when probing these features. Instead, we only need to probe up to the required scale and maximisation of the signal-to-noise should be the priority instead. In this context a survey of about 10,000 deg$^2$ should be enough. This at the same time will make it easier for the SKA1 cosmology survey to be commensal with other science cases. Finally, although we have not addressed specifically the issue of foreground contamination, it is expected that the MT technique will alleviate this problem even further since any possible residuals from the cleaning process \citep{Alonso:2014dhk} and even systematics should be uncorrelated between \textsc{Hi} IM and the photo-$z$ galaxy survey.
 
\paragraph*{Note added.}
While this paper was being completed, \cite{Alonso:2015sfa} appeared, covering a similar topic.

\paragraph*{Acknowledgements.}
We thank David Alonso, Phil Bull, and Pedro Ferreira for helpful discussions.
J.F., M.G.S. and R.M. are supported by the South African Square Kilometre Array Project and National Research Foundation. S.C. acknowledges support from the European Research Council under the EC FP7 Grant No. 280127. RM is also supported by the UK Science \&\ Technology Facilities Council Grant No. ST/K0090X/1. M.G.S. acknowledges support from FCT under grant PTDC/FIS-AST/2194/2012.


\begin{thebibliography}{37}
\expandafter\ifx\csname natexlab\endcsname\relax\def\natexlab#1{#1}\fi

\bibitem[{Abramo \& Leonard(2013)}]{Abramo:2013awa}
Abramo L.~R., Leonard K.~E., 2013, Mon. Not. Roy. Astron. Soc., 432, 318

\bibitem[{Alonso {et~al.}(2015b)Alonso, Bull, Ferreira, Maartens, \&
  Santos}]{Alonso:2015uua}
Alonso D., Bull P., Ferreira P.~G., Maartens R., Santos M.~G., 2015b,
  arXiv:1505.07596

\bibitem[{Alonso {et~al.}(2015)Alonso, Bull, Ferreira, \&
  Santos}]{Alonso:2014dhk}
Alonso D., Bull P., Ferreira P.~G., Santos M.~G., 2015, Mon. Not. Roy. Astron.
  Soc., 447, 400

\bibitem[{Alonso \& Ferreira(2015)}]{Alonso:2015sfa}
Alonso D., Ferreira P.~G., 2015, arXiv:1507.03550

\bibitem[{Amendola {et~al.}(2013)}]{Amendola:2012ys}
Amendola L., {et~al.}, 2013, Living Rev. Rel., 16, 6

\bibitem[{Bacon {et~al.}(2015)Bacon, Bridle, Abdalla, Brown, Bull,
  {et~al.}}]{Bacon:2015dqe}
Bacon D., Bridle S., Abdalla F.~B., Brown M., Bull P., {et~al.}, 2015, PoS,
  AASKA14, 145, arXiv:1501.03977

\bibitem[{Bonvin \& Durrer(2011)}]{Bonvin:2011bg}
Bonvin C., Durrer R., 2011, Phys. Rev., D84, 063505

\bibitem[{{Camera} {et~al.}(2015{\natexlab{a}}){Camera}, {Carbone}, {Fedeli},
  \& {Moscardini}}]{Camera:2014dia}
{Camera} S., {Carbone} C., {Fedeli} C., {Moscardini} L., 2015{\natexlab{a}},
  Phys. Rev., D91, 043533

\bibitem[{{Camera} {et~al.}(2015{\natexlab{b}}){Camera}, {Maartens}, \&
  {Santos}}]{Camera:2014sba}
{Camera} S., {Maartens} R., {Santos} M.~G., 2015{\natexlab{b}}, Mon. Not. Roy.
  Astron. Soc., 451, L80

\bibitem[{Camera {et~al.}(2015)Camera, Raccanelli, Bull, Bertacca, Chen,
  {et~al.}}]{Camera:2015fsa}
Camera S., Raccanelli A., Bull P., Bertacca D., Chen X., {et~al.}, 2015, PoS,
  AASKA14, 025, arXiv:1501.03851

\bibitem[{Camera {et~al.}(2012)Camera, Santos, Bacon, Jarvis, McAlpine, Norris,
  Raccanelli, \& Rottgering}]{Camera:2012ez}
Camera S., Santos M.~G., Bacon D.~J., Jarvis M.~J., McAlpine K., Norris R.~P.,
  Raccanelli A., Rottgering H., 2012, Mon. Not. Roy. Astron. Soc, 427, 2079

\bibitem[{Camera {et~al.}(2013)Camera, Santos, Ferreira, \&
  Ferramacho}]{Camera:2013kpa}
Camera S., Santos M.~G., Ferreira P.~G., Ferramacho L., 2013, Phys. Rev. Lett.,
  111, 171302

\bibitem[{{Camera} {et~al.}(2015c){Camera}, {Santos}, \&
  {Maartens}}]{Camera:2014bwa}
{Camera} S., {Santos} M.~G., {Maartens} R., 2015c, Mon. Not. Roy. Astron. Soc.,
  448, 1035

\bibitem[{Challinor \& Lewis(2011)}]{Challinor:2011bk}
Challinor A., Lewis A., 2011, Phys. Rev., D84, 043516

\bibitem[{Dalal {et~al.}(2008)Dalal, Dore, Huterer, \& Shirokov}]{Dalal:2007cu}
Dalal N., Dore O., Huterer D., Shirokov A., 2008, Phys. Rev., D77, 123514

\bibitem[{Ferramacho {et~al.}(2014)Ferramacho, Santos, Jarvis, \&
  Camera}]{Ferramacho:2014pua}
Ferramacho L.~D., Santos M.~G., Jarvis M.~J., Camera S., 2014, Mon. Not. Roy.
  Astron. Soc., 442, 2511

\bibitem[{Giannantonio {et~al.}(2012)Giannantonio, Porciani, Carron, Amara, \&
  Pillepich}]{Giannantonio:2011ya}
Giannantonio T., Porciani C., Carron J., Amara A., Pillepich A., 2012, Mon.
  Not. Roy. Astron. Soc., 422, 2854

\bibitem[{{Gong} {et~al.}(2011){Gong}, {Chen}, {Silva}, {Cooray}, \&
  {Santos}}]{2011ApJ...740L..20G}
{Gong} Y., {Chen} X., {Silva} M., {Cooray} A., {Santos} M.~G., 2011, Astrphys.
  J., 740, L20

\bibitem[{Hall {et~al.}(2013)Hall, Bonvin, \& Challinor}]{Hall:2012wd}
Hall A., Bonvin C., Challinor A., 2013, Phys. Rev., D87, 064026

\bibitem[{Hamaus {et~al.}(2011)Hamaus, Seljak, \& Desjacques}]{Hamaus:2011dq}
Hamaus N., Seljak U., Desjacques V., 2011, Phys. Rev., D84, 083509

\bibitem[{Laureijs {et~al.}(2011)}]{Laureijs:2011gra}
Laureijs R., {et~al.}, 2011, ESA-SRE, 12, arXiv:1110.3193

\bibitem[{{LSST Dark Energy Science Collaboration}(2012)}]{Abate:2012za}
{LSST Dark Energy Science Collaboration}, 2012, arXiv:1211.0310

\bibitem[{Ma {et~al.}(2005)Ma, Hu, \& Huterer}]{Ma:2005rc}
Ma Z.-M., Hu W., Huterer D., 2005, Astrophys. J., 636, 21

\bibitem[{Maartens {et~al.}(2015)Maartens, Abdalla, Jarvis, \&
  Santos}]{Maartens:2015mra}
Maartens R., Abdalla F.~B., Jarvis M., Santos M.~G., 2015, PoS, AASKA14, 016, arXiv:1501.04076

\bibitem[{Matarrese \& Verde(2008)}]{Matarrese:2008nc}
Matarrese S., Verde L., 2008, Astrophys. J., 677, L77

\bibitem[{McDonald \& Seljak(2009)}]{McDonald:2008sh}
McDonald P., Seljak U., 2009, JCAP, 0910, 007

\bibitem[{{Montanari} \& {Durrer}(2015)}]{Montanari:2015rga}
{Montanari} F., {Durrer} R., 2015, arXiv:1506.01369

\bibitem[{Namikawa {et~al.}(2011)Namikawa, Okamura, \&
  Taruya}]{Namikawa:2011yr}
Namikawa T., Okamura T., Taruya A., 2011, Phys. Rev., D83, 123514

\bibitem[{{Planck Collaboration} {et~al.}(2015){Planck Collaboration}, {Ade},
  {Aghanim}, {Arnaud}, {Arroja}, {Ashdown}, {Aumont}, {Baccigalupi},
  {Ballardini}, {Banday}, \& et~al.}]{Ade:2015ava}
{Planck Collaboration}, {Ade} P.~A.~R., {Aghanim} N., {Arnaud} M., {Arroja} F.,
  {Ashdown} M., {Aumont} J., {Baccigalupi} C., {Ballardini} M., {Banday} A.~J.,
  et~al., 2015, arXiv:1502.01592

\bibitem[{{Raccanelli} {et~al.}(2015){Raccanelli}, {Montanari}, {Bertacca},
  {Dor{\'e}}, \& {Durrer}}]{Raccanelli:2015vla}
{Raccanelli} A., {Montanari} F., {Bertacca} D., {Dor{\'e}} O., {Durrer} R.,
  2015, arXiv:1505.06179

\bibitem[{Santos {et~al.}(2015)Santos, Bull, Alonso, Camera, Ferreira,
  {et~al.}}]{Santos:2015bsa}
Santos M., Bull P., Alonso D., Camera S., Ferreira P., {et~al.}, 2015, PoS,
  AASKA14, 019, arXiv:1501.03989

\bibitem[{Seljak(2009)}]{Seljak:2008xr}
Seljak U., 2009, Phys. Rev. Lett., 102, 021302

\bibitem[{Seljak {et~al.}(2009)Seljak, Hamaus, \& Desjacques}]{Seljak:2009af}
Seljak U., Hamaus N., Desjacques V., 2009, Phys. Rev. Lett., 103, 091303

\bibitem[{Tegmark {et~al.}(1997)Tegmark, Taylor, \& Heavens}]{Tegmark:1996bz}
Tegmark M., Taylor A., Heavens A., 1997, Astrophys. J., 480, 22

\bibitem[{Yamauchi {et~al.}(2014)Yamauchi, Takahashi, \&
  Oguri}]{Yamauchi:2014ioa}
Yamauchi D., Takahashi K., Oguri M., 2014, Phys. Rev., D90, 083520

\bibitem[{Yoo(2010)}]{Yoo:2010ni}
Yoo J., 2010, Phys. Rev., D82, 083508

\bibitem[{Yoo {et~al.}(2012)Yoo, Hamaus, Seljak, \& Zaldarriaga}]{Yoo:2012se}
Yoo J., Hamaus N., Seljak U., Zaldarriaga M., 2012, Phys. Rev., D86, 063514

\end{thebibliography}
\end{document}